\documentclass[
	a4paper,
	aps,
	prb,
	twocolumn,
	longbibliography,
	preprintnumbers,
	amsmath,
	amssymb,
	floatfix,
	superscriptaddress,
	nofootinbib11,
	]{revtex4-2}

\usepackage{graphicx}% Include figure files
\usepackage{dcolumn}% Align table columns on decimal point
\usepackage{bm}% bold math
\usepackage{verbatim}
\usepackage{tensor}
\usepackage{bbm}
\usepackage{tabularx}
\usepackage{slashed}
\usepackage{cancel}
\usepackage[colorlinks=true,
						urlcolor=magenta,
						linkcolor=blue,
						citecolor=blue,
						filecolor=black,
						menucolor=blue,
						anchorcolor=red,
						runcolor=cyan
						]{hyperref}
\usepackage{etoolbox}
\usepackage{orcidlink}

\def\be{\begin{equation}}
\def\ee{\end{equation}}
\def\bea{\begin{eqnarray}}
\def\eea{\end{eqnarray}}
\def\ba#1\ea{\begin{align}#1\end{align}}
\def\bg#1\eg{\begin{gather}#1\end{gather}}
\def\bm#1\em{\begin{multline}#1\end{multline}}
\def\bmd#1\emd{\begin{multlined}#1\end{multlined}}

\newcommand{\sgn}{{\rm{sgn}}}

\begin{document}

% \preprint{APS/123-QED}

\title{Reduce dimensional quantum criticality for Non-Fermi liquids}
%\title{Reduced dimension quantum field theory for Non-Fermi liquid theory\\
%Non-Fermi liquid fixed point for reduce dimensional quantum critical metals}
%\title{Reduced dimension quantum field theory for Non-Fermi liquid theory}% Force line breaks with \\

\author{Phumudzo T. Rabambi\orcidlink{}}
\email{teflon.ac.za@gmail.com}
\affiliation{\small \it Department of Physics, University of Witswatersrand,	Wits, 2050, South Africa }

\author{Mario Sol\'{i}s\orcidlink{0000-0001-6139-075X}}
\email{mario.solis-benites@pwr.edu.pl}
\affiliation{\small \it Institute of Theoretical Physics, Wroc\l{}aw University of Science and Technology, 50-370 Wroc\l{}aw, Poland}

\begin{abstract}
We present a reduced dimension theoretical framework for studying quantum field theories at finite density, providing a tractable model for investigating non-Fermi liquid (NFL) behavior near quantum phase transitions. Our approach departs from the standard paradigm by placing bosons and fermions in different spatial dimensions: bosonic fields reside in a $(d+1)$-dimensional bulk, while fermionic fields are confined on a $d$-dimensional boundary. This dimensional separation significantly simplifies the renormalization group (RG) analysis of gapless boson-fermion coupling. We demonstrate that the tree-level boson exchange contributions, which typically exhibit logarithmic divergences, become finite in our reduced-dimension scheme. Furthermore, the $\log^2$ and $\log^3$ divergences that characterize 1-loop four-fermion interactions in conventional treatments are reduced to logarithmic divergences within this framework, substantially improving the convergence properties of the perturbative expansion and allowing controlled theoretical analysis of NFL physics.
\end{abstract}


\maketitle

%\tableofcontents

%%%%%%%%%%%%%%%%%%%%%%%%%%%%%%%%%%%%%%%%%%%%%%%%
%%%%%%%%%%%%%%%%%%%%%%%%%%%%%%%%%%%%%%%%%%%%%%%%
%%%%%%%%%%%%%%%%%%%%%%%%%%%%%%%%%%%%%%%%%%%%%%%%
%%%%%%%%%%%%%%%%%%%%%%%%%%%%%%%%%%%%%%%%%%%%%%%%
\section{Introduction}
{A promising  framework for understanding strongly correlated electronic system is finite density quantum field theory (QFT). One of the applications of QFT at finite density is  quantum phase transition of metallic system. A metal is described by the  Landau Fermi liquid theory \cite{Landau:1956zuh}, it stands as one of the most successful theoretical frameworks for explaining the thermodynamic and transport properties across a wide spectrum of energies. However, a significant portion of strongly correlated materials exhibit a breakdown of Fermi liquid theory when they undergo continuous quantum phase transitions in the vicinity of quantum critical points (QCPs). In this vicinity, Non-Fermi liquid behavior emerges, characterized by anomalous thermodynamic and transport properties \cite{Sachdev_2011,RevModPhys.79.1015, RevModPhys.73.797}. Understanding NFL dynamics remains an open and challenging problem; therefore, developing a general consistent and controllable theoretical framework to capture the physics of NFL is of significant interest in condensed matter physics.

The experimental evidence regarding NFL behavior is that near QCP, quantum fluctuations of the order parameter are responsible for transforming a Fermi liquid into a Non-Fermi liquid while simultaneously driving an enhanced of superconductivity dynamics. The measurements on those strongly correlated materials, such as cuprates \cite{broun2008lies}, iron pnictides \cite{shibauchi2014quantum},  organic \cite{dressel2011quantum}, heavy fermions \cite{stewart1984heavy} and so on, shown something in common: the presence of a superconducting dome near the QCP. In order to reveal the exact location of this QCP is required to suppress superconductivity by tuning the control parameter to its critical value e.g. by strong magnetic field.}

The theoretical approach to understanding NFL typically employs effective low-energy field theories \cite{SHANKAR1991530, polchinski1999effectivefieldtheoryfermi, RevModPhys.66.129}. The simple model of fermionic quasi-particles near the Fermi surface couple to gapless bosonic modes \cite{Sachdev_2011,RevModPhys.79.1015, RevModPhys.73.797, Shuster_2000, PhysRevB.72.174520, PhysRevB.80.184518, 8481185917bc4a5cb66147ad833131cd, PhysRevD.60.114033, PhysRevLett.110.127001, PhysRevB.88.245106, PhysRevB.80.165102}, representing collective fluctuations of order parameters. However, treated at equal footing the fermionic and bosonic modes can presents some technical and fundamental challenges to the renormalization group of the theory.

The first technical challenge is the appearance of the tree-level boson exchange diagram; responsible to the enhancement of BCS instability, is logarithmic divergent. Inducing the 1-loop four-fermion interactions diagrams to appear with $\log^2$ ($\log^3$) divergences instead of the standard log divergences, which is a second technical challenge \cite{Mahajan2013QuantumCM, Torroba:2014gqa, PhysRevB.89.165114, PhysRevB.92.045118, fitzpatrick2014slow, Fitzpatrick_2015}. The boson exchange diagram is proportional to the bosonic propagator, in $d=3$ it is schematically to
\be
g_0^2\int d^3q\,\frac{1}{q_0^2 + q^2} \sim g_0^2\log(q)\,,
\ee
producing a logarithm divergent contribution.

The appearance of tree-level $\log$ divergences and 1-loop $\log^2$ ($\log^3$) divergences can invalidate the fundamental notion of fixed points if these divergences are not carefully removed in the RG analysis, therefore the inclusion of Feynman diagrams that are not 1PI (1-Particle Irreducible) is permitted in the RG analysis in order to fully remove these divergences from the theory, which complicates the RG analysis.

In this work, we present a novel theoretical framework based on reduced dimensional separation, where the free boson fields reside in $(d+1)$-dimensional bulk space, while the fermion fields are confined to a $d$-dimensional boundary (or brane in high-energy physics context), and all the relevant interactions in the theory are on the $d$-dimensional boundary. This setup, inspired by holographic approaches and reduced QED models \cite{giombi2019models,Son_2007,Teber_2018}, offers several key advantages that are elaborated in our main results. 

Within our reduced dimension framework, we show that such tree-level $\log$ divergence and 1-loop $\log^2$ ($\log^3$) divergences are absent, therefore only 1PI Feynman diagrams are permitted in the RG analysis, which significantly simplifies the RG structure and analysis in finite density systems for NFL models. 

The paper is organized as follows. In Section \ref{sec.reduced_dim_model}, we demonstrate why both the $\log^2$ and $\log^3$ strong divergences at 1-loop and the $\log$ divergences at tree level are absent in the reduced dimension framework. Our first main result is at equation (\ref{tree_reduced}), we find that the tree-level boson exchange propagator is finite and not $\log$ divergent, which then leads to the 1-loop four-fermion diagrams having the standard log divergences. The second main result is showed at equation (\ref{ld2}), we confirm that since the boson tree-level exchange propagator is finite, this subsequently leads to the 1-loop four-fermion interaction Feynman diagrams having the standard log divergences, and the $\log^2$ ($\log^3$) divergences are absent. Contrary to conventional approaches, here the tree-level $\log$ and 1-loop $\log^2$ ($\log^3$) divergences are being tamed by reorganizing the degrees of freedom of the theory, through adding an additional dimension for the boson field and confining the interactions of the theory on the $d$-dimension boundary.

In Section \ref{sec.RG_flow}, we renormalize the theory by evaluating the relevant 1-loop diagrams, followed by deriving the relevant beta functions at this order, for simplicity we turn-off the 4-fermion coupling; we leave the analysis of it in upcoming work. The main results obtained in Sec. \ref{sec.reduced_dim_model} simplify the RG analysis for the theory since the tree-level boson exchange logarithm divergence is absent and also the 1-loop four-fermion interactions $\log^2$ ($\log^3$) divergences are absent. Consequently, in our RG analysis, we find straightfoward RG equations with tractable divergences and fixed points. Our results establish a lot of new perspective to construct NFL models.

We end the paper with a discussion and future direction in Sec. \ref{sec.discussion}. We also present detailed calculations on the appendix.

%%%%%%%%%%%%%%%%%%%%%%%%%%%
%%%%%%%%%%%%%%%%%%%%%%%%%%%
%%%%%%%%%%%%%%%%%%%%%%%%%%%
%%%%%%%%%%%%%%%%%%%%%%%%%%%
\section{Reduced Dimension Model}\label{sec.reduced_dim_model}
The idea of reduced-dimension QFT is adopted from previous studies of QFT in reduced dimensions \cite{giombi2019models,Son_2007,Teber_2018}. We begin by proposing an effective quantum field theory at low energy to study NFL theories using reduced dimensions. The theory consists of an undamped critical (massless) boson interacting with a Fermi surface of spinless fermions via the Yukawa coupling. We focus on the case $d$-spatial dimensions, where the theory is weakly coupled, enabling a controlled perturbative expansion in $d=3$. In $(d+1)$ space and time dimensions the bare Euclidean action is
\be
\mathcal{S} = \int d\tau d^dx\,\mathcal{L}_{\phi} +\int d\tau  d^{d-1}x\,\mathcal{L}_{\psi} +\int dxd^{d-1}\mathcal{L}_I,  
\ee
where the lagrangians are
\bea
\mathcal{L}_{\phi}&=&(\partial_\tau \phi)^2 + c^2 (\nabla\phi)^2 \,, \\
\mathcal{L}_{\psi} &=&  \psi^\dagger(i\partial_\tau + \epsilon_F (i\nabla) - \mu_F) \psi\,,
\eea
and the interaction lagrangian is
\be\label{eq.int_action}
\mathcal{L}_I = g\psi^\dagger\psi\phi + \lambda_\psi (\psi^\dag \psi)^2\,. 
\ee
The fermionic and scalar field (the order parameter of the theory) are expressed with $\psi$ and $\phi$ respectively. The $\epsilon_F$ is the quasiparticle (fermion field) energy and $\mu_F$ the chemical potential of the fermion field. Here the scalar field $\phi$ lives in $(d+1$)-dimensions in the bulk with the mass set to criticality, and the fermion field lives in $d$-dimensions on the boundary (or Brane \cite{giombi2019models,Gorbar_2001,Kaplan_2009}), and the Yukawa interactions are on the boundary. Here the boson momenta scale towards the origin
\be
p_0^\prime=e^tp_0\,,\quad \vec{p}\,^\prime = e^t\vec{p}\,
\ee
and the fermion momenta scale towards the Fermi surface
\begin{eqnarray}
p_0^\prime = e^tp_0,\quad p_\perp^\prime = e^t p_\perp\,,
\end{eqnarray}
where $p_{\perp}$ is perpendicular to the Fermi surface. To keep the action classically invariant the fields scale as follows
\be
\phi^\prime(p^\prime)=e^{-\frac{d+3}{2}t}\phi(p),\quad \psi^\prime(p^\prime)=e^{-\frac{3}{2}t}\psi(p)\,,
\ee
and
\be
\lambda_{\psi}^\prime=\lambda_{\psi}\,,\quad g^\prime = e^{\frac{d-3}{2}t}g\,,
\ee
for the interactions couplings. We just consider the marginal interactions.

The main motivation for adopting reduced-dimension theory here is that it greatly simplifies the RG flow analysis of a massless boson coupled to the fermion Fermi surface at finite density. As established from conventional approaches in \cite{PhysRevB.92.045118, Fitzpatrick_2015}, four-fermion interactions start to run at tree level due to the appearance of log divergence from the tree level boson-exchange, and the 1-loop Feynman diagrams for four-fermion interactions results in $\log^2$ and $\log^3$ divergences, leading to non-1PI (none 1-Particle Irreducible) diagrams required to renormalize the theory. The four-fermion interactions in the reduced-dimension theory in equation \eqref{eq.int_action} offer less complex RG flow analysis with manageable degrees of divergence compared to those found in \cite{PhysRevB.92.045118, Fitzpatrick_2015}. Here the strong $\log^2$ and $\log^3$ divergences are being tamed by reorganising the degrees of freedom of the theory, through adding an additional dimension for the boson field and confining the interactions of the theory on the d-dimension boundary. Below we demonstrate the contrast in degrees of divergence obtained between both frameworks. 

\begin{figure}[h]
\includegraphics[width=0.45\textwidth]{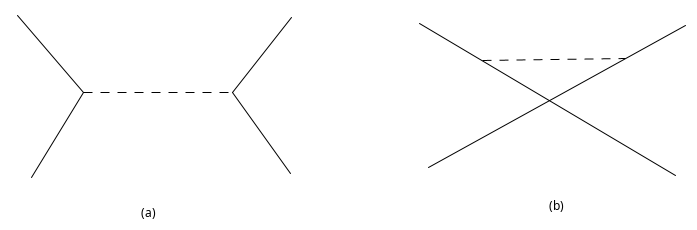}
\caption{Tree-level boson exchange process diagram and 1-loop four-Fermi diagram}
\label{fig:first}
\end{figure}

We first show the degrees of divergence obtained from a conventional framework adopted in \cite{PhysRevB.92.045118, Fitzpatrick_2015} and compare these with the reduced-dimension framework results obtained here. The tree-level diagram shown in Fig.~\ref{fig:first}(a) represents a boson-mediated scattering process in the Cooper channel, and its propagator is given as follows
\begin{equation}
V_{k,k'} = \frac{g^2}{(k_0 - k_0')^2 + c^2 |\vec{k} - \vec{k}'|^2}, 
\end{equation}
where $k$ and $k'$ are the external fermion momenta and $c=1$. It has been shown in \cite{PhysRevB.92.045118} that there is a logarithmic divergence embedded in this tree-level propagator, which is revealed by decomposing the propagator into angular momentum harmonics
\begin{equation}
V_L = \frac{1}{2} \int_{-1}^{1} d(\cos \theta) V_{k,k'} P_L(\cos \theta), 
\end{equation}
where $\theta$ is the angle between $k$ and $k'$. For simplicity, the case $L = 0$ was used, and the integral was treated in a Wilsonian approach by integrating out only the fast bosonic modes over the momentum shell $(\Lambda - d\Lambda) < q < \Lambda$. The results show the tree-level term to be logarithmically divergent
\begin{equation}
\delta V_0 \simeq -\frac{g^2}{2k_F^2}\frac{d\Lambda}{\Lambda}.
\end{equation}
Evaluating the same tree-level boson exchange propagator in the reduced-dimension framework, where the boson-mediated process scattering propagator in reduced-dimension takes the form,
\begin{equation}
V_{k,k^\prime} = \frac{1}{2}\frac{g^2}{((k_0 - k_0')^2 + |\vec{k} - \vec{k}'|^2)^{\frac{1}{2}}}, 
\end{equation}
and adopting the same evaluation process, in reduced-dimensions we obtain a finite tree-level term
\begin{equation}
\label{tree_reduced}
\delta V_0 = -\frac{g^2}{2k_F^2}d\Lambda.
\end{equation}
Now the tree-level boson exchange being finite, this has consequences for the 1-loop four-fermion couplings. The 1-loop four-fermion couplings are now reduced from $\log^2$ and $\log^3$ divergences to just $\log$ divergences. It is shown in \cite{PhysRevB.92.045118, Fitzpatrick_2015} that the effect contributing to the $\log^2$ and $\log^3$ divergences emanate from the tree-level boson exchange process being logarithmically divergent. 

The evaluation of one of the 1-loop four-fermion coupling Feynman diagrams shown in Fig.~\ref{fig:first}(b), following the conventional approach \cite{PhysRevB.92.045118},  reveals the following $\log^2$ divergence
\begin{widetext} 
\begin{eqnarray}
\Gamma^4(k)_{d+1} &=& g^2 \lambda_{BCS} \int \frac{dq_0 d^3 q}{(2\pi)^4} D(k - q)G(q)G(-q)\,,\nonumber\\
&=& g^2 \lambda_{BCS} \int \frac{dq_0 d^3 q}{(2\pi)^4} \frac{1}{(k_0 - q_0)^2 + 2k_F^2 (1 - \cos \theta)} \frac{1}{iq_0 - vq_{\bot}} \frac{1}{-iq_0 - vq_{\bot}} \,, \nonumber\\
&=& -\frac{\lambda_{BCS}g^2 c^2 k_F^2}{2k_F^2} \log(\Lambda) \log\left( \frac{\Lambda}{\Lambda \cos \phi}\right) + \cdots \,,
\end{eqnarray}
where $D(k-q)$ and $G(q)$ are the boson and fermion propagators, respectively.  The coupling $\lambda_{BCS}$ denotes the $\lambda_\psi$ coupling after restrict it to the BCS channel. Now, evaluating the same 1-loop four-fermion coupling diagram in the reduced-dimension framework at the boundary, we obtain the normal expected logarithmic divergences at 1-loop,
\begin{eqnarray}
\label{ld2}
 \Gamma^4(k)_{d}&=& g^2 \lambda_{BCS} \int \frac{dq_0 d^2 q}{(2\pi)^3} D(k - q)G(q)G(-q)\,,\nonumber\\
&=& \frac{g^2 \lambda_{BCS}}{2} \int \frac{dq_0 d^2 q}{(2\pi)^3}\frac{1}{((k_0 - q_0)^2 + 2k_F^2 (1 - \cos \theta))^\frac{1}{2}} \frac{1}{iq_0 - vq_{\bot}} \frac{1}{-iq_0 - vq_{\bot}}\,, \nonumber\\
&=& \frac{g^2 k_F^2 \lambda_{BCS}}{2} \int \frac{dq_0 dq_{\bot} d(\cos \theta)}{(2\pi)^2} \frac{1}{2}\frac{1}{((k_0 - q_0)^2 + 2k_F^2 (1 - \cos \theta))^{\frac{1}{2}}} \frac{1}{q_0^2 + v^2q^2_{\bot}}\,, \nonumber \\
&=&\frac{g^2 k_F \lambda_{BCS}}{(2\pi)v}\int_0^\infty\frac{dq_0}{q_0}\,,
\end{eqnarray}
\end{widetext}
where the last line on eq. \eqref{ld2} shows that the integral diverges logarithmically both at the infrared ($q_0\rightarrow 0$) and ultraviolet ($q_0\rightarrow \infty$) limits. The results obtained above in equation \eqref{ld2} demonstrate the absence of $\log^2$ and $\log^3$ divergences at 1-loop in the reduced-dimension framework theory, because there is no tree-level running from the four-fermion coupling arising from the boson exchange process.

%%%%%%%%%%%%%%%%%%%%%%%%%%%
%%%%%%%%%%%%%%%%%%%%%%%%%%%
%%%%%%%%%%%%%%%%%%%%%%%%%%%
%%%%%%%%%%%%%%%%%%%%%%%%%%%
\section{Renormalization Group Flow}\label{sec.RG_flow}
\begin{figure}[h!]
\includegraphics[width=0.45\textwidth]{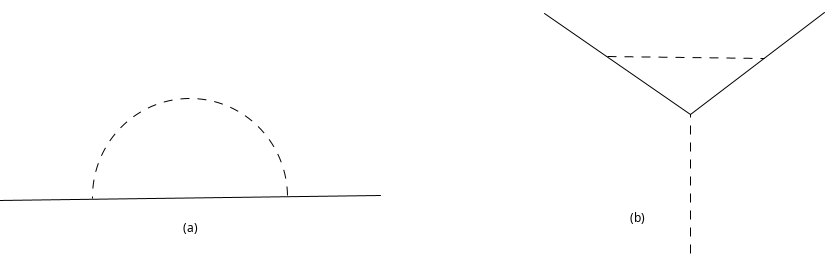}
\caption{1-loop fermion self energy and 1-loop Yukawa coupling}
\label{fig:second}
\end{figure}

In this section, we will perform a perturbative $\epsilon$ expansion of the theory in $d=3-\epsilon$ to obtain the 1-loop quantum corrections of the theory. As a starting point we set the four-fermion interactions to $\lambda_{\psi}=0$ just to ease the computation, and their inclusion will be considered in a future work. The action becomes
\bea
S &=&\int  d\tau d^2x\, \left((\psi^\dagger_0(\partial_{\tau}- \epsilon_F +\mu_F)\psi_0 + g_0\phi_0\psi^\dagger_0\psi_0\right)+  \nonumber \\
&& \int d\tau d^3x\,\left((\partial_{\tau}\phi_0)^2+(\vec{\nabla}\phi_0)^2\right)\,,
\eea
and the relevant 1-loop diagrams are shown in Figure \ref{fig:second}. The 1-loop quantum corrections from these diagrams will have poles at $\epsilon\rightarrow 0$, and counterterms will be introduced to subtract these poles to obtain physical results. The dependence of counterterms on $\epsilon$ will then be used to obtain the relevant beta functions of the theory. We start by writing down the action in terms of bare parameters (fields and couplings) which are denoted by the subscript $0$, and these bare parameters are expressed in terms of physical parameters and counterterms as follows
\be
\label{bare}
\psi_0=\sqrt{Z_{\psi}}\psi\,, \quad v_0=Z_{v}v\,,\quad g_0=\mu^{\frac{\epsilon}{2}}\frac{Z_g}{Z_{\psi}}\,,\quad \phi_0=\sqrt{Z_{\phi}}\phi \,,
\ee
where $\phi$ and $\psi$ are physical fields and $\mu$ is an arbitrary scale. Noting that $Z_{\phi}=1$, since the scalar field is in the bulk and free.
From here we define the counterterms,
\be
Z_{\psi}=1+\delta_{\psi}\,,\quad \delta_v=Z_{v}v_0-v\,,\quad g\delta_g = g_0Z_{\psi}-g\,.
\ee
Differentiating both sides of the bare field equations in eq. \eqref{bare} with respect to the scale $\mu$, we obtain the fermion anomalous dimension, the beta function for the velocity and coupling $g$. These equantions are
\begin{eqnarray}
\label{flow}
\gamma_{\psi}&=&\frac{1}{2}\mu\frac{d\delta_{\psi}}{d\mu} \,, \nonumber \\
\beta_{v}&=&2\gamma_{\psi}v-\mu\frac{d\delta_{v}}{d\mu}\,, \nonumber \\
\beta_g &=& \bigg(-\frac{\epsilon}{2}+\gamma_{\psi}-\mu\frac{d\delta_g}{d\mu}\bigg)g\,.
\end{eqnarray}

Since $Z_{\phi}=1$, the boson field anomalous dimension $\gamma_{\phi}=0$, as the scalar field does not receive any quantum corrections. After computing the relevant 1-loop integrals shown in the appendix \ref{app.detail_comp}, the counterterms $\delta_{\psi}$, $\delta_g$ and $\delta_v$ are extracted as
\begin{eqnarray}
\label{3-point}
&&\delta_{\psi}=-\frac{1}{2}\bigg(\frac{g^2}{2\pi^2\sqrt{4\pi}(1+|v|)}\frac{1}{\epsilon}\bigg),\\\nonumber
&&\delta_v=\frac{1}{2}\bigg(\frac{g^2\text{sgn}(v)}{2\pi^2\sqrt{4\pi}(1+|v|)}\frac{1}{\epsilon}\bigg),\\\nonumber
&&\delta_g=-\frac{1}{2}\bigg(\frac{2g^3\pi^2\csc(\frac{\pi\epsilon}{2})}{(2\pi)^3\sqrt{4\pi}}\frac{(iq_0+\text{sgn}(v)q_{\bot})}{(1+|v|)(iq_0-vq_{\bot})}\bigg).\\\nonumber
\end{eqnarray}
Adopting the renormalization strategy used in \cite{Fitzpatrick_2015}, where $\delta_g$ can be further renormalized by letting $q_0=x\mu$, $q_{\bot}=\mu$ and taking $x\rightarrow \infty$ as follows
\begin{eqnarray}
\delta_g &=& -\frac{1}{2}\bigg(\frac{g^3}{2\pi^2\sqrt{4\pi}}\frac{\big(i+\frac{\sgn(v)}{x}\big)}{\big(1+|v|\big)\big(i-\frac{v}{x}\big)}\frac{1}{\epsilon}\bigg)\,,\nonumber \\
&=& -\frac{1}{2}\bigg(\frac{g^3}{2\pi^2\sqrt{4\pi}}\frac{1}{(1+|v|)}\frac{1}{\epsilon}\bigg)\,.
\end{eqnarray}
\noindent Substituting these terms back into the flow equations \eqref{flow}, we obtain
\begin{eqnarray}
\gamma_{\psi}=&&\frac{1}{2}\frac{g^2}{4\pi^2\sqrt{4\pi}(1+|v|)},\\\nonumber
\beta_{v}=&& 2\bigg(\frac{1}{2}\frac{g^2}{4\pi^2\sqrt{4\pi}(1+|v|)}\bigg)v+\frac{1}{2}\frac{g^2\text{sgn}(v)}{2\pi^2\sqrt{4\pi}(1+|v|)},\\\nonumber
=&&\frac{1}{2}\frac{g^2}{4\pi^2\sqrt{\pi}}\text{sgn}(v),\\\nonumber
\beta_g=&&-\frac{\epsilon}{4}g.
\end{eqnarray}

\noindent The results for $\gamma_{\psi}$  and $\beta_v$ are similar to the results obtained in \cite{PhysRevB.92.045118, Fitzpatrick_2015}, but with an extra factor of $\frac{1}{2}\frac{1}{\sqrt{4\pi}}$ due to the theory being in reduced dimensions. At $\epsilon=0$, the Yukawa coupling is marginal, and $\beta_v$ has an attractive fixed point where $v\rightarrow 0$. The flow of the velocity $v$ reaches its limiting scale at a finite scale
\begin{eqnarray}
\mu_{v=0}=e^{-\frac{2\pi^2\sqrt{4\pi}}{g_0^2}}\Lambda,
\end{eqnarray}
where $g_0$ and $v_0$ are the values at the scale $\Lambda$. At $v\rightarrow 0$, the non-Fermi liquid has an anomalous dimension
\be
\gamma_{\psi}=\frac{g^2}{4\pi^2\sqrt{4\pi}}\,.
\ee
We can appreciate this a positive anomalous dimension and it usually tends to make the 4-fermion interaction more irrelevant. We will study this situation in a future.

%%%%%%%%%%%%%%%%%%%%%%%%%
%%%%%%%%%%%%%%%%%%%%%%%%%
%%%%%%%%%%%%%%%%%%%%%%%%%
%%%%%%%%%%%%%%%%%%%%%%%%%
\section{Discussion and future directions}\label{sec.discussion}
In this paper we have presented a reduced-dimension field theoretical framework to study a simple theory of massless bosons coupled to the fermion Fermi surface, where the Yukawa interactions is on the boundary which is one dimension less, and the considered renormalization group flow is also on the boundary. 

The tree-level boson exchange process term is finite, which subsequently leads to the absence of $\log^2$ and $\log^3$ divergences in the four-fermion coupling at 1-loop. The absence of such strong divergences is due to the degrees of freedom of the theory being reorganized through adding an additional dimension for the boson field and confining the interactions of the theory on the d-dimension boundary. The presence of $\log^2$ and $\log^3$ divergences at 1-loop arising from conventional setups \cite{PhysRevB.92.045118, Fitzpatrick_2015} can invalidate the notion of fixed points if the RG analysis doesn't rigorously enable the full cancellation of these strong divergences. The absence of such $\log^2$ and $\log^3$ divergences in our framework will pose no threat to the notion of valid fixed points when four-fermion interactions are included later in the RG analysis. The demonstration that 1-loop four-fermion processes exhibit only logarithmic divergences, as shown in equation \eqref{ld2}, suggests that the perturbative expansion in the reduced-dimension framework may remain well-controlled to higher orders. This is in stark contrast to conventional approaches where the proliferation of $\log^2$ and $\log^3$ divergences in 1-loop signal a potential breakdown of the notion of fixed points. The anomalous dimension we computed, $\gamma_{\psi}=\frac{g^2}{4\pi^2\sqrt{4\pi}}$, provides a concrete realization of NFL behaviour within a controlled theoretical framework, which provides a significant step forward in understanding quantum critical metals using tractable models. The anomalous dimension solution while similar in structure to previous results, carries the additional factor of $\frac{1}{2}\frac{1}{\sqrt{\pi}}$ which emanates from the reduced-dimensional nature of the theory and may lead to different critical exponents compared to standard treatments.

The practical implications of our results extend beyond mere technical improvements in calculation control. By eliminating the most problematic divergences that plague conventional approaches, our framework opens new avenues for systematic study of NFL behavior near quantum critical points. The finite tree-level boson exchange we obtained in equation \eqref{tree_reduced} represents a fundamental departure from standard treatments, where such processes typically introduce strong logarithmic divergences that cascade into higher-order corrections. This finiteness stems directly from the dimensional separation between bulk bosons and boundary fermions, which effectively decouples the most troublesome aspects of boson-fermion interactions while preserving the essential physics of criticality. This is a significant improvement in comparison to the other conventional finite density field theory models for NFL where the bosons and fermions are treated on equal footing. The only drawback with this framework is that UV-IR mixing would still be present, which is due to the different kinematics structure of the fermions and bosons: the low energies for the fermions is near a Fermi surface, whereas for the boson the low energies are at low momentum near a point. As already observed in \cite{Torroba:2014gqa}, this difference in kinematics structure will induce the 1-loop Yukawa coupling 3-point function to have an $\epsilon$ pole which has a nonlocal and singular factor of momenta $(p_0, p_\perp)$ as shown in the last line of equation \eqref{3-point} in Section \ref{sec.RG_flow}. The results from the last line of equation \eqref{3-point} are a contribution from both the UV region and IR region- integrating the low energy degrees of freedom (IR region) near the Fermi surface yields a singular denominator factor in momenta ($p_0, p_{\perp}$), whereas the integration of high energy degrees of freedom (UV region) yields the $\epsilon$ pole.

In Section \ref{sec.RG_flow}, the term four-fermion coupling was removed from the RG analysis for the $\epsilon$ expansion in $d=3-\epsilon$; the next immediate task for future work will be to include the four-fermion coupling in the full RG analysis. Do an analysis of 2-loop and higher-loop divergences for the four-fermion coupling, since normally at 2-loops we anticipate controllable $\log^2$ divergences within this framework. This analysis will be crucial for determining whether our dimensional separation technique merely postpones problematic uncontrollable strong divergences to higher orders or fundamentally resolves them. Stated differently, this will help verify whether there is any proliferation of unconventional strong divergences at 2-loops and higher loops, which would indicate limitations of our approach. This is essential for a complete understanding of the flow structure of the theory and for identifying additional fixed points that may emerge from the interplay between Yukawa and four-fermion interactions responsible for BCS (Bardeen–Cooper–Schrieffer) and forward-scattering interactions. 

The results presented here, in our reduced dimensional framework, opens a plethora of opportunities to future work. An interesting thing to explore is the inclusion of four-fermion coupling. In previous study \cite{Raghu:2015sna}, the presence of tree-level logarithmic divergence boson exchange propagator is associated with the enhancement of BKT transitions in conventional NFL theoretical models \cite{Kaplan:2009kr, Raghu:2015sna}, it contributes an extra enhancement term in the beta function for the BCS coupling. In reduced dimension framework, the boson exchange propagator is finite, therefore we anticipate the enhancement of BKT transition will be absent from this framework of NFL models. Beyond the inclusion of four-fermions coupling and higher loops we intend to extend the work by finding exact relations for the fermion self-energy and Yukawa coupling non-perturbatively \cite{Huh_2008}, then proceed to calculate the exact critical exponents. Other application of our framework is to study of Landau parameters and the effects the Friedel oscillations, as in \cite{Fitzpatrick_2015} in conventional approach; we would like to explore all these ideas in a future work.

%%%%%%%%%%%%%%%%%%%%%%%%%
%%%%%%%%%%%%%%%%%%%%%%%%%
%%%%%%%%%%%%%%%%%%%%%%%%%
%%%%%%%%%%%%%%%%%%%%%%%%%
\section{Acknowledgments}
The authors would like to thank Robert De Mello Koch, Jaco Van Zyl, Ki-Seok Kim, Anatolii Kotikov, Paweł Jakubczyk, Jeremias Aguilera-Damia and Gonzalo Torroba for useful discussions and comments on the paper. M.S. was supported in part by the Polish National Science Centre (NCN) Sonata Bis grant 2019/34/E/ST3/00405.
%%%%%%%%%%%%%%%%%%%%%%%%%
%%%%%%%%%%%%%%%%%%%%%%%%%
%%%%%%%%%%%%%%%%%%%%%%%%%
%%%%%%%%%%%%%%%%%%%%%%%%%
\begin{widetext}
\appendix
\section{Detailed computation}\label{app.detail_comp}

In this appendix, we provide detailed calculations that support the main results presented in the body of the paper. The calculations are organized into three subsections that demonstrate the key technical advantages of our reduced-dimension framework.

The first subsection establishes the fundamental difference between conventional and reduced-dimension treatments by explicitly computing the tree-level boson exchange propagator and the 1-loop four-fermion coupling corrections. These calculations directly verify our central claim that the dimensional separation eliminates the problematic $\log^2$ and $\log^3$ divergences that plague standard approaches. The second and third subsections present the detailed evaluation of the 1-loop Feynman integrals required for the renormalization group analysis in Section~3, including the fermion self-energy and vertex corrections that determine the anomalous dimensions and beta functions of the theory.

Throughout these calculations, we work in dimensional regularization with $d = 3 - \epsilon$ and employ standard techniques including Feynman parameterization and momentum shifting. The key technical innovation that enables our results is the modified boson propagator structure taking the form $\sim (k^2)^{-1/2}$ rather than the conventional form $\sim (k^2)^{-1}$, which arises naturally from the dimensional separation between bulk bosons and boundary fermions.
%%%%%%%%%%%%%%%%%%%%%%%%%
%%%%%%%%%%%%%%%%%%%%%%%%%
\subsection{Absence of strong divergences in reduced-dimension framework}

\noindent This subsection demonstrates the crucial advantage of our reduced-dimension approach by explicitly comparing the degree of divergences in tree-level and 1-loop processes with those found in conventional treatments. The calculations below directly support the claims made in Section~2 regarding the finiteness of tree-level boson exchange and the absence of $\log^2$ and $\log^3$ divergences at 1-loop.

We will first begin by deriving the boson propagator in reduced dimension at (2+1)-dimension. Here the boson propagator is the integral of the (3+1)-dimension propagator,
\begin{eqnarray}
\bar{D}(k)=\frac{1}{k_0^2+k_1^2+k_2^2+k_3^2}\,,
\end{eqnarray}
integrated over the momentum component perpendicular to the (2+1)-dimension plane, which will be the $k_3$ plane,
\begin{eqnarray}
D(k)=&&\int\frac{dk_3}{(2\pi)}\frac{1}{k_0^2+k_1^2+k_2^2+k_3^2}\,,\nonumber\\
    =&&\frac{1}{2}\frac{1}{(k_0^2+k_1^2+k_2^2)^{\frac{1}{2}}}\,.
\end{eqnarray}
Below is the detailed tree-level evaluation of the boson propagator in reduced dimensions:
\begin{eqnarray}
\delta V_0&=&\frac{g^2}{4}\int \frac{d(\cos\theta)}{((k_0-k_0^\prime)^2+|\vec{k}-\vec{k}^\prime|^2)^{\frac{1}{2}}}\,,
\end{eqnarray}
we define the momentum transfer $q$ as follows,
\begin{eqnarray}
    &&q^2=|\vec{k}-\vec{k}^\prime|\simeq 2k_f^2(1-\cos\theta)\,,\\
    &&\Rightarrow d(q^2)=-2k_F^2d(\cos\theta)\,,
\end{eqnarray}
Therefore,
\begin{eqnarray}
\delta V_0&=&-\frac{g^2}{8k_F^2}\int\frac{d(q^2)}{((k_0-k_0^\prime)^2+q^2)^{\frac{1}{2}}}\,.
\end{eqnarray}
Now, let
\begin{eqnarray}
&&y=\sqrt{(k_0-k_0^{\prime})^2+q^2}\,,\\
&&\Rightarrow dy = \frac{1}{2}\frac{d(q^2)}{y}\,,
\end{eqnarray}
then
\begin{eqnarray}
V_0 &=&-\frac{g^2}{4k_F^2}\int\frac{d(q^2)}{((k_0-k_0^\prime)^2+q^2)^{\frac{1}{2}}}\,,\\
&=&-\frac{g^2}{4k_F^2}\int \frac{ydy}{y}\,,\\
&=&-\frac{g^2}{4k_F^2}\int_{\sqrt{(k_0-k_0^{\prime})^2+(\Lambda-d\Lambda)^2}}^{\sqrt{(k_0-k_0^{\prime})^2+\Lambda^2}}dy \,,\\
&=&-\frac{g^2}{4k_F^2}d\Lambda\,.
\end{eqnarray}
This integral evaluates to a finite result, in stark contrast to the logarithmic divergence found in conventional treatments. Now, the corresponding computation for the 1-loop four-Fermi vertex $\Gamma^4(k)_{d+2}$ in reduced dimensions is
\begin{eqnarray}
\Gamma^4(k)_{d}=&&g^2\lambda_{BCS}\int\frac{dq_0d^2q}{(2\pi)^3}D(k-q)G(q)G(-q)\,,\\
=&&\frac{g^2\lambda_{BCS}}{2}\int\frac{dq_0d^2q}{(2\pi)^3}\frac{1}{((k_0-q_0)^2+2k_F^2(1-\cos\theta))^{\frac{1}{2}}}\frac{1}{iq_0-vq_{\bot}}\frac{1}{-iq_0-vq_{\bot}}\,,
\end{eqnarray}
where $D(k-q)$ is the boson propagator, $G(q)$ and $G(-q)$ are the fermion propagators,
\begin{eqnarray}
    D(k-q)&&=\frac{1}{2}\frac{1}{((k_0-q_0)^2+2k_F^2(1-\cos\theta))^{1/2}}\nonumber\,,\\
    G(q) &&= \frac{1}{iq_0-vq_{\bot}}\,,
\end{eqnarray}
and $q_{\bot}$ is the perpendicular distance from the Fermi surface to the fermion. We set the measure to $d^2q=2\pi k_F^2dq_{\bot}d(\cos\theta)$ and let $y=vq_{\bot}$.
\begin{eqnarray}
\Gamma^4(k)_{d}=&&\frac{g^2k_F^2\lambda_{BCS}}{2v}\int\frac{dq_0dyd(\cos\theta)}{(2\pi)^2}\frac{1}{((k_0-q_0)^2+2k_F^2(1-\cos\theta))^{\frac{1}{2}}}\frac{1}{q_0^2+y^2}\,.
\end{eqnarray}
The integral over $\cos\theta$ yields,
\begin{eqnarray}
\int_{-1}^1d(\cos\theta)\frac{1}{((k_0-q_0)^2+2k_F^2(1-\cos\theta))^{\frac{1}{2}}}=\frac{1}{k_F^2}\bigg(\sqrt{4k_F^2+(k_0-q_0)^2}-(k_0-q_0)\bigg)
\simeq \frac{1}{2k_F^2}(2k_F)=\frac{1}{k_F}\,,
\end{eqnarray}
\noindent Crucially, the integration over $\cos\theta$ (Fermi surface) doesn't produce the logarithmic term that appears in conventional calculations. This is the key difference that eliminates the cascade of strong divergences. The remaining integral is evaluated as follows:
\begin{eqnarray}
\Gamma^4(k)_{d}=&&\frac{g^2k_F\lambda_{BCS}}{2(2\pi)^2v}\int_{-\infty}^{\infty}dq_0\int_{-\infty}^{\infty}dy\frac{1}{q_0^2+y^2}\,,\\
=&&\frac{g^2k_F\lambda_{BCS}}{(2\pi)v}\int_0^\infty\frac{dq_0}{q_0}\,,
\end{eqnarray}
The final expression shows only a simple logarithmic divergences, confirming the absence of the problematic $\log^2$ terms found in conventional approaches.
%%%%%%%%%%%%%%%%%%%%%%%%%
%%%%%%%%%%%%%%%%%%%%%%%%%
\subsection{Fermion self-energy loop calculations}
This subsection presents the detailed calculation of the 1-loop fermion self-energy correction shown in Fig.~\ref{fig:second}(a). This diagram contributes to the fermion anomalous dimension $\gamma_\psi$ and is essential for determining the renormalization group flow of the theory. The calculation employs dimensional regularization in $d = 3 - \epsilon$ dimensions and demonstrates the controlled nature of divergences in reduced-dimension framework. The 1-loop fermion self-energy integral is
\begin{eqnarray}
I_{int}&=&\frac{1}{2}\int \frac{dp_{0}dp_{\bot} d^{d-2}p_{\|}}{(2\pi)^{d}}\frac{1}{(p_0^2+p^2_{\bot}+p^2_{\|})^{\frac{1}{2}}}\frac{1}{i(p_0+k_0)-v(p_{\bot}+k_{\bot})}\,.
\end{eqnarray}
Using Feynman parameterization:
\begin{eqnarray}
\frac{1}{A_1^{m_1}A_2^{m_2}\cdots A_{n}^{m_n}}=\int_0^1dx_1\cdots dx_n\delta(\sum_ix_i-1)\frac{\prod x_i^{m_i-1}}{(\sum x_iA_i)^{\sum m_i}}\frac{\Gamma(\sum m_i)}{\prod\Gamma(m_i)} \,,
\end{eqnarray}
\noindent the integral can be re-written as follows,
\begin{eqnarray}
I_{int}=\frac{1}{2}\frac{\Gamma(\frac{3}{2})}{\Gamma(\frac{1}{2})}\int \frac{dp_{0}dp_{\bot} d^{d-2}p_{\|}}{(2\pi)^{d}}\int_0^1 dx dy\frac{\delta(x+y-1)x^{\frac{1}{2}-1}(i(p_0+k_0)+v(p_{\bot}+k_{\bot}))}{[(p_0^2+p^2_{\bot}+p^2_{\|})x+((p_0+k_0)^2+v^2(p_{\bot}+k_{\bot})^2)y]^{1+\frac{1}{2}}},
\end{eqnarray}
\noindent Performing parameter shifts $p_0\rightarrow p_0-k_0y$ and $p_{\bot}\rightarrow p_{\bot}-\frac{v^2k_{\bot}y}{v^2y+x}$ to complete the square:
\begin{eqnarray}
p^2_0 &=& p_0^2-2p_0k_0y+k_0^2y^2\,,\\
p_{\bot}^2 &=& p_{\bot}^2-\frac{2v^2p_{\bot}k_{\bot}y}{v^2y+x}+\frac{v^4k_{\bot}^2y^2}{(v^2y+x)^2}\,.
\end{eqnarray}
\noindent Applying the shifts and absorbing the $\int dy$ integral, the integral becomes:
\begin{eqnarray}
I_{int}&=&\frac{1}{2}\frac{\Gamma(\frac{3}{2})}{\Gamma(\frac{1}{2})}\int \frac{dp_{0}dp_{\bot} d^{d-2}p_{\|}}{(2\pi)^{d}}\int_0^1 dx\frac{x^{-\frac{1}{2}}(i(k_0-\delta_0)+v(k_{\bot}-\delta_{\bot}))}{[p_0^2+(v^2-(v^2-1)x)p_{\bot}^2+p_{\|}^2x+k_0^2x(1-x)+v^2k_{\bot}^2(1-x)x]^{\frac{3}{2}}}\,,
\end{eqnarray}
where $\delta_0=k_0(1-x)$, $\delta_\bot = \frac{v^2k_\bot(1-x)}{v^2(1-x)+x}$, and we have dropped terms odd in $p$, since the integrals evaluate to zero. Using the standard integral formula:
\begin{eqnarray}
I_n &=&\int dp_0dp_{\bot}d^{d-2}p_{\|}\frac{1}{(Ap_0^2+Bp_{\bot}^2+Cp_{\|}^2+\bigtriangleup)^n}\,,\\
&=&\frac{\pi^{\frac{d}{2}}\Gamma(\frac{2n-d}{2})}{\Gamma(n)}\frac{1}{\sqrt{ABC^{d-2}\bigtriangleup^{2n-d}}}\,,
\end{eqnarray}
then the integral above at $d=3-\epsilon$ evaluates to
\begin{eqnarray}
I_{int}&=&\frac{1}{2}\frac{\pi^{2}\csc\frac{\pi\epsilon}{2}}{(2\pi)^3}\bigg(\frac{2(ik_0+\text{sgn}(v)k_{\bot})}{(1+|v|)}\bigg)\,, \nonumber \\
&=&\frac{\pi}{(2\pi)^3}\bigg(\frac{2(ik_0+\text{sgn}(v)k_{\bot})}{(1+|v|)}\bigg)\frac{1}{\epsilon}+\mathcal{O}(\epsilon)\,.
\end{eqnarray}
%%%%%%%%%%%%%%%%%%%%%%%%%
%%%%%%%%%%%%%%%%%%%%%%%%%
\subsection{1-loop vertex calculations for Yukawa interactions}
This subsection presents the calculation of the 1-loop vertex correction to the Yukawa coupling shown in Fig.~\ref{fig:second}(b). This three-point function determines the renormalization of the Yukawa coupling constant and contributes to the beta function $\beta_g$. The calculation demonstrates another instance where the reduced-dimension framework leads to well-controlled divergences compared to conventional treatments. The three-point integral is:
\begin{eqnarray}
I_3 &=& \frac{1}{2}\int \frac{dp_{0}dp_{\bot}d^{d-2}p_{\|}}{(2\pi)^{d}}\frac{1}{(p_0^2+p_{\bot}^2+p_{\|}^2)^{\frac{1}{2}}}\frac{1}{i(p_0+q_0)-v(p_{\bot}+q_{\bot})}\frac{1}{ip_0-vp_{\bot}}\,,\\
&=&\frac{1}{2}\int \frac{dp_{0}dp_{\bot}d^{d-2}p_{\|}}{(2\pi)^{d}}\int dxdydz\frac{\delta(x+y+z-1)z^{1-\frac{1}{2}}[i(p_0+q_0)+v(p_{\bot}+q_{\bot})][ip_0+vp_{\bot}]}{(p_0^2+(v^2(x+y)+z)p_{\bot}^2+zp_{\|}^2+(2p_0q_0+q_0^2+v^2(2p_{\bot}q_{\bot}+q_{\bot}^2))y)^{2+\frac{1}{2}}}\,,
\end{eqnarray}
making the following transformations
\begin{eqnarray}
&&p_{0}\rightarrow p_{0}-q_{0}y\\
&&p_{\bot}\rightarrow p_{\bot}-\frac{v^2q_{\bot}y}{v^2(x+y)+z},
\end{eqnarray}
the integral becomes
\be
I_{3}=\int \frac{dp_{0}dp_{\bot}d^{d-1}p_{\|}}{(2\pi)^{d+1}}\int dxdydz\frac{\delta(x+y+z-1)z^{-\frac{1}{2}}[i(p_0+q_0-\delta_0)+v(p_{\bot}+q_{\bot}-\delta_{\bot})][i(p_0-\delta_0)+v(p_{\bot}-\delta_{\bot})]}{(p_0^2+(v^2(x+y)+z)p_{\bot}^2+p_{\|}^2z+q_0^2y(1-y)+v^2q_{\bot}^2y(v^2x+z))^{2+\frac{1}{2}}}\,,
\ee
\noindent where $\delta_0=q_0y$, $\delta_\bot = \frac{v^2q_{\bot}y}{v^2(x+y)+z}$. Dropping numerator terms odd in $p_{0}$ and $p_{\bot}$ since they evaluate to zero, and using the following integrals
\begin{eqnarray}
I_{n1} =&&\int dp_0dp_{\bot}d^{d-2}p_{\|}\frac{1}{(Ap_0^2+Bp_{\bot}^2+Cp_{\|}^2+\bigtriangleup)^n}=\frac{\pi^{\frac{d+1}{2}}\Gamma(\frac{2n-d-1}{2})}{\Gamma(n)}\frac{1}{\sqrt{ABC^{d-1}\bigtriangleup^{2n-d}}}\,,\\
I_{n2}=&&\int dp_0dp_{\bot}d^{d-2}p_{\|}\frac{p_0^2}{(Ap_0^2+Bp_{\bot}^2+Cp_{\|}^2+\bigtriangleup)^n}=\frac{\pi^{\frac{d+1}{2}}\Gamma(\frac{2n-d-3}{2})}{2\Gamma(n)}\frac{1}{\sqrt{A^3BC^{d-1}\bigtriangleup^{2n-d-2}}}\,,\\
I_{n3}=&&\int dp_0dp_{\bot}d^{d-2}p_{\|}\frac{p_{\bot}^2}{(Ap_0^2+Bp_{\bot}^2+Cp_{\|}^2+\bigtriangleup)^n}=\frac{\pi^{\frac{d+1}{2}}\Gamma(\frac{2n-d-3}{2})}{2\Gamma(n)}\frac{1}{\sqrt{AB^3C^{d-1}\bigtriangleup^{2n-d-2}}}\,.
\end{eqnarray}
\noindent At $d=3-\epsilon$ the integral evaluates to
\begin{eqnarray}
I_{3}=&&\frac{1}{2} \frac{\pi^2\csc(\frac{\pi\epsilon}{2})}{2(2\pi)^3}\frac{\Gamma(\frac{1}{2})}{\Gamma(\frac{5}{2})}\frac{4(iq_0+\text{sign}(v)q_{\bot})}{(|v|+1)(iq_0-vq_{\bot})}\,.
\end{eqnarray}

\end{widetext}
%\nocite{*}

\bibliography{apssamp}{}% Produces the bibliography via BibTeX.
\bibliographystyle{utphys}

\end{document}